\documentclass[a4paper,fleqn,usenatbib]{mnras}
\usepackage{graphicx}	% Including figure files
\usepackage{amsmath}	% Advanced maths commands
\usepackage{amssymb}	% Extra maths symbols
\usepackage{multicol}        % Multi-column entries in tables
\usepackage{bm}		% Bold maths symbols, including upright Greek
\usepackage{pdflscape}	% Landscape pages

\usepackage{multirow}
\usepackage{textcomp}
\usepackage{amssymb}

\usepackage{hyperref}

\usepackage[T1]{fontenc}
\usepackage{ae,aecompl}

\usepackage[utf8]{inputenc} 

\newcommand{\cxo}{{\em Chandra}}

\newcommand{\e}{{\em Einstein}}
\newcommand{\xmm}{{\em XMM--Newton}}

\newcommand{\swift}{{\em Swift}}

\newcommand{\nus}{{\em NuSTAR}}

\def\nh {$N_{H}$}

\def\rchisq {$\chi_{\nu}^{2}$}
\def\lum {erg\,s$^{-1}$}
\def\flux {erg\,s$^{-1}$cm$^{-2}$}

\def\cm {cm$^{-2}$}
\def\arcsec{$^{\prime\prime}$}

\def\srclong{1E\,161348--5055}
\def\src{1E\,1613}
\def\snr{RCW\,103}

\title[\nus\ and \swift\ observations of \srclong]{Gazing at the ultra-slow magnetar in \snr\ with \nus\ and \swift}

\author[A. Borghese et al.]{
A. Borghese,$^{1}$\thanks{E-mail: a.borghese@uva.nl} F. Coti Zelati,$^{2,3}$ P. Esposito,$^{1}$ N. Rea,$^{1,2,3}$ A. De Luca,$^{4}$ M. Bachetti,$^{5}$ \and G.L. Israel,$^{6}$ R. Perna$^{7}$ and J. A. Pons$^{8}$
\\
$^{1}$Anton Pannekoek Institute for Astronomy, University of Amsterdam, Postbus 94249, NL--1090 GE Amsterdam, The Netherlands\\
$^{2}$Institute of Space Sciences (ICE, CSIC), Campus UAB, Carrer de Can Magrans s/n, 08193 Barcelona, Spain\\
$^{3}$Institut d’Estudis Espacials de Catalunya (IEEC), 08034 Barcelona, Spain\\
$^{4}$INAF-Istituto di Astrofisica Spaziale e Fisica Cosmica, Milano, via E. Bassini 15, I-20133 Milano, Italy\\
$^{5}$INAF-Osservatorio Astronomico di Cagliari, via della Scienza 5, I-09047 Selargius (CA), Italy\\ 
$^{6}$INAF-Osservatorio Astronomico di Roma, via Frascati 33, I-00040, Monteporzio Catone (RM), Italy\\
$^{7}$Department of Physics and Astronomy, Stony Brook University, Stony Brook, NY, 11794, USA\\
$^{8}$Departament de F\'isica Aplicada, Universitat d'Alacant, Ap. Correus 99, 03080 Alacant, Spain
}

\date{Accepted XXX. Received YYY; in original form ZZZ}

\pubyear{2018}

\begin{document}
\label{firstpage}
\pagerange{\pageref{firstpage}--\pageref{lastpage}}
\maketitle

\begin{abstract}

We report on a new \nus\ observation and on the ongoing \swift\ XRT monitoring
campaign of the peculiar source \srclong, located at the centre of the
supernova remnant \snr, which is recovering from its last outburst in June
2016. The X-ray spectrum at the epoch of the \nus\ observation
can be described by either two absorbed blackbodies ($kT_{BB_1}$
$\sim$ 0.5~keV, $kT_{BB_2}$ $\sim$ 1.2~keV) or an absorbed blackbody
plus a power law ($kT_{BB_1}$ $\sim$ 0.6~keV, $\Gamma$ $\sim$
3.9). The observed flux was $\sim$ 9 $\times$ 10$^{-12}$~\flux, $\sim$
3 times lower than what observed at the outburst onset, but 
about one order of magnitude higher than the historical quiescent level. 
A periodic modulation was detected at the known 6.67~hr periodicity.
The spectral decomposition and evolution along the outburst decay are
consistent with \srclong\ being a magnetar, the slowest ever detected.

 \end{abstract}


\section{Introduction}

The discovery of \srclong\ (\src\ hereafter) dates back to the year 1980, when
the \e\ observatory detected a point-like X-ray source laying close to
the geometrical centre of the young ($\sim$ 2~kyr;
\citealt{1997PASP..109..990C}) supernova remnant (SNR)
\snr\ \citep{1980ApJ...239L.107T}.
\src\ was identified as the first radio-quiet, cooling isolated
neutron star (NS) in a SNR, characterized by soft thermal X-ray
emission, lack of counterparts at other wavelengths and no detected
pulsations. Since then, a few objects with similar properties had been
observed in other SNRs and grouped in the class of `central compact
objects' (CCOs; see \citealt{1742-6596-932-1-012006} for a
review). To date, this class includes a dozen of sources, including the 
recently CCO identified outside the Milky Way in the SNR 1E\,0102.2-7219 \citep{2018arXiv180301006V}.
Pulsations have been detected for three of them only, unveiling fast
periods ($P$ $<$ 0.5~s) and a weak dipolar magnetic field at the
surface ($B_{dip}$ $\lesssim$ 10$^{11}$~G).

During the last two decades, however, some remarkable features were
observed in \src, separating it from the CCOs. Firstly, unlike the other CCOs 
that have generally a steady emission, \src\ shows a strong flux variability on a month/year
time scale. In 1999 it experienced an outburst that yielded an increase
in flux by a factor of $\sim$ 100 \citep{2000HEAD....5.3211G}. Secondly,
thanks to a long ($\sim$ 90~ks) \xmm\ observation that caught the
source in a low state, a periodicity of 6.67 $\pm$ 0.03~hr was
revealed with a strong, almost sinusoidal modulation
\citep{2006Sci...313..814D}. Although the 6.67~hr periodicity could be
recognized in all the long-enough data sets, the corresponding pulse
profile changed according to the source flux level: from sine-like
shape when the source is in a low state (observed 0.5 -- 8~keV flux
$\sim$ 10$^{-12}$~\flux) to more complex, multi-peaked configurations in high state
($\sim$ 10$^{-11}$~\flux). The long-term variability and the long
periodicity have contributed to build a unique
phenomenology in the NS scenario. Based on these characteristics, two
main interpretations were put forward: \src\ could be either a
low-mass X-ray binary (LMXB) in a SNR \citep{2009A&A...506.1297B} or a
peculiar isolated object with a rotational period of 6.67~hr
\citep{2007ApJ...666L..81L,2006Sci...313..814D}. In the former
hypothesis, the 6.67~hr periodicity would be the orbital period of the
system. To explain the intriguing behaviour, an unusual double
accretion mechanism, involving wind and accretion disk, was suggested
\citep{2006Sci...313..814D}. On the other hand, considering the
isolated-object hypothesis, the magnetar scenario would naturally
account for the flux and pulse shape variations. Magnetars, isolated
NSs powered by ultra-strong magnetic fields (typically $B$ $\sim$
10$^{13-14}$~G), are characterized by high-energy flaring events
(e.g. short bursts and outbursts) and spin periods in the range 0.3 --
12~s (see \citealt{2017ARA&A..55..261K} for a review). An
efficient braking mechanism needs to be invoked in order to slow down
\src\ from a fast birth period ($<$ 0.5~s) to such a long rotational period 
in $\sim$ 2~kyr. Most models consider a propeller interaction with a fall-back disk that can
provide an additional spin-down torque besides that due to dipole
radiation. \citet{2017MNRAS.464L..65H} predict a remnant disk of
10$^{-9}$~M$_\odot$ around a rapidly rotating NS that initially is in
an ejector phase, and after hundreds of years its rotation becomes
slow enough to allow the onset of a propeller phase, during which
strong slow-down torques cause an increase of the spin period. In
order to reproduce the observational properties (the period of 6.67~hr
and the young age), \src\ is found to have a slightly higher magnetic
field than known magnetars, $\sim$ 5 $\times$ 10$^{15}$~G (for the
surface dipolar component). \citet{2016ApJ...833..265T} considered a
similar model but obtained a larger disk mass of $\sim$
10$^{-5}$~M$_\odot$ and a comparable magnetic field strength.

Recently, a new event shed light on the nature of this source: on 2016 June 22 a short magnetar-like burst of hard X-rays from the direction of \snr\ triggered the {\em Neil Gehrels Swift Observatory} (\swift) Burst Alert Telescope (BAT). 
The \swift\ X-ray Telescope (XRT) was observing the source till $\sim$
20 minutes before the BAT trigger and detected a 1 -- 10~keV flux
enhancement of a factor of $\sim$ 100 with respect to the quiescent
level ($\sim$ 10$^{-12}$~\flux), measured one month before \citep{2016MNRAS.463.2394D, 2016ApJ...828L..13R}. Target of
Opportunity observations were performed a few days later with \cxo\ and
the {\em Nuclear Spectroscopic Telescope Array} (\nus). \nus\ detected
a hard X-ray, non-thermal component up to $\sim$ 30~keV, modelled by a
power law with photon index $\Gamma$ = 1.20 $\pm$ 0.25 and pulsed till
$\sim$ 20~keV; on the other hand, the soft \cxo\ spectrum is well described by two
blackbodies \citep{2016ApJ...828L..13R}. The light curves exhibit two
peaks per cycle, in contrast with the sinusoidal shape observed during
the quiescent state \citep{2011MNRAS.418..170E}. This outburst
prompted new searches for an infrared counterpart. The {\em Hubble
  Space Telescope} observed \src\ twice after the burst
\citep{2017ApJ...841...11T}. The images disclosed a new object in the
\cxo\ position ellipse, not detected in previous observations acquired
in 2002. The counterpart properties rule out the binary scenario: the
2002 upper limits and the new detection would be consistent with a
compact hydrogen atmosphere white dwarf as a companion; the high
surface gravity of this object implies a star radius $\leq$
5000~km. This radius is smaller than the orbital radius (1.7 $\times$
10$^6$~km) and the corresponding Roche lobe radius for a binary system
of a white dwarf and a NS orbiting with a 6.67~hr period. As a
consequence, accretion cannot occur and power the X-ray flux
variations. Furthermore, the X-ray to infrared fluence ratio of $\sim$
10$^5$ is compatible with the one measured for magnetars
(\citealt{2011AdSpR..47.1281M}; see also the McGill Magnetar
Catalog\footnote{
  \url{http://www.physics.mcgill.ca/~pulsar/magnetar/main.html}}). However,
it is not clear whether the emission comes from the NS magnetosphere
or from a fall-back disk. The properties of the outbursts, the discovery of
a non-thermal hard component in the spectrum and an infrared
counterpart, the pulse profile variability in time and energy are all
hints that point towards the magnetar scenario.

Here we report on a new \nus\ observation of \src, performed 345 days after its last outburst, and a long-term \swift\ XRT monitoring campaign, focusing on the observations carried out from the onset of the 2016 outburst till October 2017. Data reduction is described in Section \ref{data}; analysis and results follow in Section \ref{analysis}. Finally, our findings are discussed in Section \ref{disc}.

\section{Observations and data reduction}
\label{data}

\src\ was observed simultaneously with \nus\ and \swift\ on 2017 June 2. In addition, a \swift\ XRT monitoring campaign has been carried out since 2006 with monthly observations. Data reduction was performed using tools incorporated in {\sc heasoft} (version 6.22). We report uncertainties at 1$\sigma$ confidence level for a single parameter of interest, unless otherwise noted. Throughout this work we use the most accurate position of \src\ derived by \citet{ 2008ApJ...682.1185D} from \cxo\ data, $\rm RA = 16^h17^m36^s.23$ and $\rm Dec = -51^\circ02'24\farcs6$ (J2000.0), and a distance of 3.3~kpc \citep{1975A&A....45..239C}.

\subsection{{\em NuSTAR}}

\nus\ is the first focusing hard X-ray space observatory, launched on 2012 June 13 \citep{2013ApJ...770..103H}. It consists of two co-aligned optics focused onto two focal planes (FPMA and FPMB), observing the sky in the energy range from 3 to 79~keV. Its effective collecting area peaks at $\sim$ 900~cm$^2$ around 10~keV, adding up the two modules. It achieves an energy resolution of 400~eV at 10~keV and an angular resolution of 18\arcsec\ full width at half maximum.

\nus\ observed \src\ between 2017 June 2 and 3, for a total dead-time
corrected on-source exposure time of 59.7~ks and 57.9~ks for FPMA and
FPMB, respectively (observation ID: 30301017002\footnote{Both modules
  are contaminated by stray light (photons that are not reflected by the focusing mirror); the pattern is more evident on a
  chip different from that of the source.}). We reprocessed the data
using the calibration files {\sc caldb} version 20171002. We ran the
tool {\sc nupipeline} to generate cleaned event lists and remove the
passages of the observatory through the South Atlantic
Anomaly. Afterwards, we referred the photon arrival times to the
Solar system barycentre reference frame by means of the {\sc barycorr}
task, using the Solar system ephemeris DE200, the \cxo\ CCO position and version 75 of the \nus\ clock
correction file that accounts for drifts caused by temperature
variations. We extracted the source photons from a circular region of
radius 40\arcsec\ and the background counts through two different
regions, a 60\arcsec- and 100\arcsec-radius circles, far from the
source location, but on the same chip. Due to the {\sc nuproducts}
script, we produced light curves, background-subtracted spectra and
the corresponding redistribution matrices and ancillary response files
separately for both focal plane modules. The energy channels outside
the range 3 -- 79~keV were flagged as bad.

\begin{figure*}
\begin{center}
\includegraphics[scale=0.5,trim={0 0cm 0 0cm},clip=true]{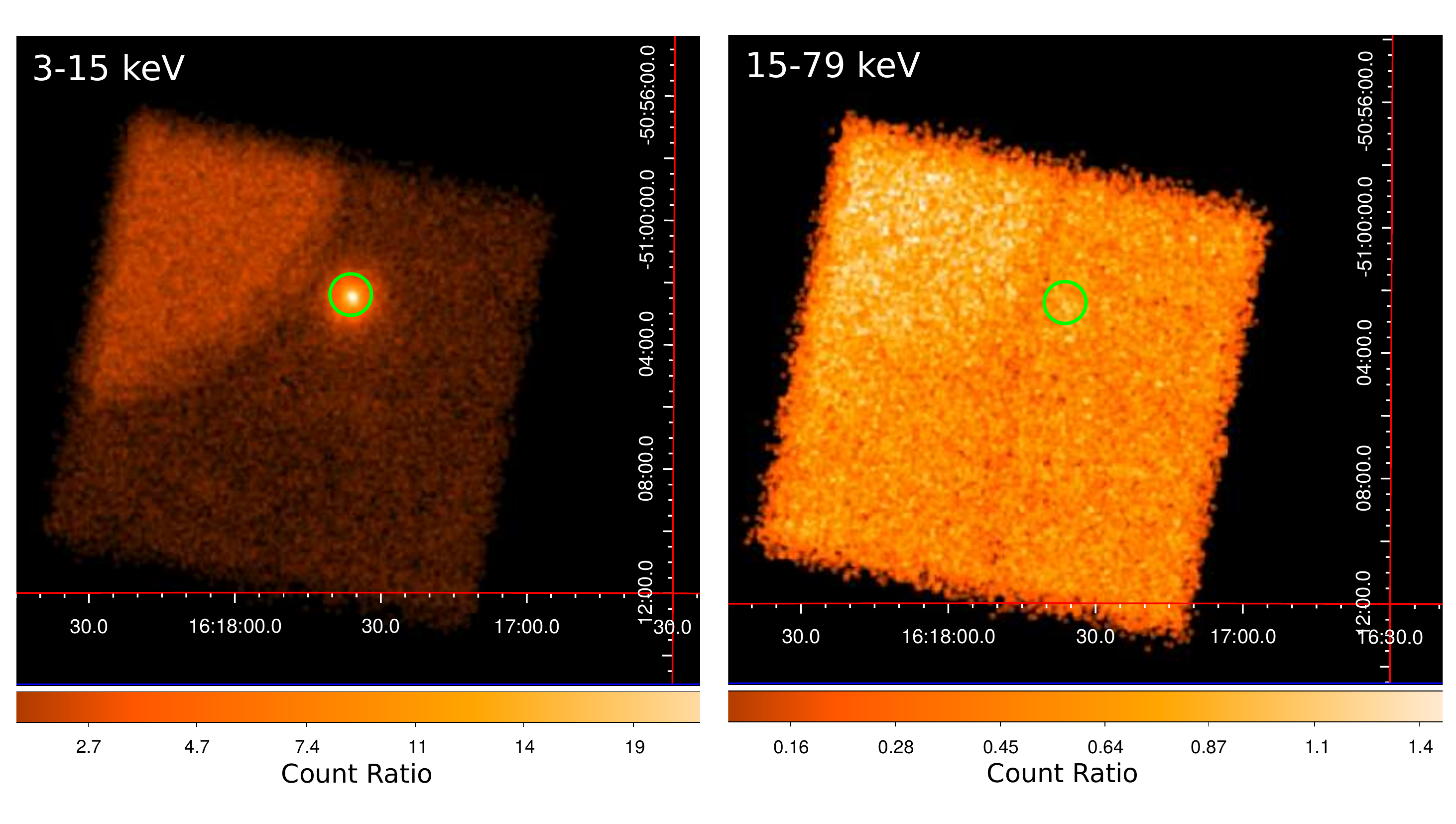}
\vspace{-5pt} 
\caption{\nus\ FPMA+FPMB exposure corrected-images for observation 30301017002 in the 3 -- 15~keV (left panel) and 15 -- 79~keV (right panel) energy bands. North is up, East to the left. Both images were smoothed with a Gaussian filter with a kernel of 3 pixels. The green circle with a radius of 40\arcsec\ represents the extraction region we used to collect the source photons. Stray light (photons that are not reflected by the focusing mirror) is present in the North-East corner of the image.}
\label{fig:fovnus}
%\vskip -0.1truecm
\end{center}
\end{figure*}

We studied the signal-to-noise ratio (S/N) of the point source with {\sc XIMAGE} in energy bands of increasing width (3 -- 10~keV, 3 -- 11~keV, etc.). We detected \src\ with a S/N $\sim$ 64 in the 3 -- 15~keV energy range. Above 15~keV the S/N ceased to increase, implying that the source emission starts to be comparable to the background level at this energy. The source net count rates were 0.057 $\pm$ 0.001 ~counts s$^{-1}$ and 0.051 $\pm$ 0.001~counts s$^{-1}$ for FPMA and FPMB, respectively, in the 3 -- 15~keV energy range. Figure \ref{fig:fovnus} shows the exposure corrected-images in two different energy bands (3 -- 15~keV and 15 -- 79~keV) for the combined event file of FPMA and FPMB data.

%%%%%%%%%%%%%%%%%%%%%%%%%%%%%%%%%%%%%%%%%%%%%%%%%%%%%%%%%%%%%%%%%%%%%%%%%%%%%%%%%%%%%%%%%%%
%%%%%%%%%%%%%%%%%%%%%%%%%%%%%%%%%%%%%%%%%%%%%%%%%%%%%%%%%%%%%%%%%%%%%%%%%%%%%%%%%%%%%%%%%%

\subsection{{\em Swift}}

\swift\ XRT is a focusing X-ray telescope with a 110~cm$^2$ effective area at 1.5~keV and a 24-arcmin field of view, sensitive to photons in the energy range 0.2 -- 10~keV \citep{2005SSRv..120..165B}. Its CCD detector can be operated in two main readout modes: photon counting (PC) and windowed timing (WT). PC mode retains full imaging resolution with a time resolution of $\sim$ 2.51~s; while in WT mode only one dimensional information is preserved with a 1.77~ms time resolution.
 
\src\ was observed by XRT 52 times since the onset of its last
outburst on 2016 June 22 and until 2017 October 16\footnote{On 2017
  June 2, simultaneously to \nus, XRT observed \src\ for a total
  dead-time corrected on-source time of 6.4~ks in WT mode and 160~s in
  PC mode (observation ID: 00088149001). To study the SNR
  contamination in the WT mode data, we extracted its spectrum from
  different regions and found that the SNR emission is still bright
  over $\sim$ 2~keV, making it challenging to discern properly the source
  contribution. Therefore, we preferred not to include this
  observation in our analysis.}. The single exposure lengths varied
between $\sim$ 0.4 and $\sim$ 6.4~ks, summing up to a total of $\sim$
41.8~ks of dead-time corrected on-source exposure time. The monitoring
campaign was rather intensive until the end of 2016 October (about 2
-- 4 observations per week), when the source entered a solar
constraint. Observations resumed in 2017 mid-January, and were
subsequently performed on a monthly cadence as part of our approved
monitoring program (PI: De Luca). The XRT was configured in PC mode in
all observations.

We reprocessed the data with standard screening criteria, generated exposure maps with the task {\sc xrtpipeline} and referred the photon arrival times to the Solar system reference frame using the {\sc barycorr} tool and the best source position. Source and background counts were extracted adopting the same regions as in \citet[see also \citealt{2011MNRAS.418..170E}]{2016ApJ...828L..13R}: a circle of radius 10 pixels for the source photons and an annulus of inner and outer radii of 10 and 20 pixels for the background counts (1 XRT pixel corresponds to about 2.36 arcsec). We extracted the corresponding spectra using {\sc xselect} and created auxiliary files with the {\sc xrtmkarf} tool. The response matrice version `20130101v014' available in the XRT calibration database was assigned to each spectrum.

\section{Analysis and results}
\label{analysis}

\subsection{Timing analysis}
\label{timing}
\begin{figure}
\begin{center}
\includegraphics[scale=0.33]{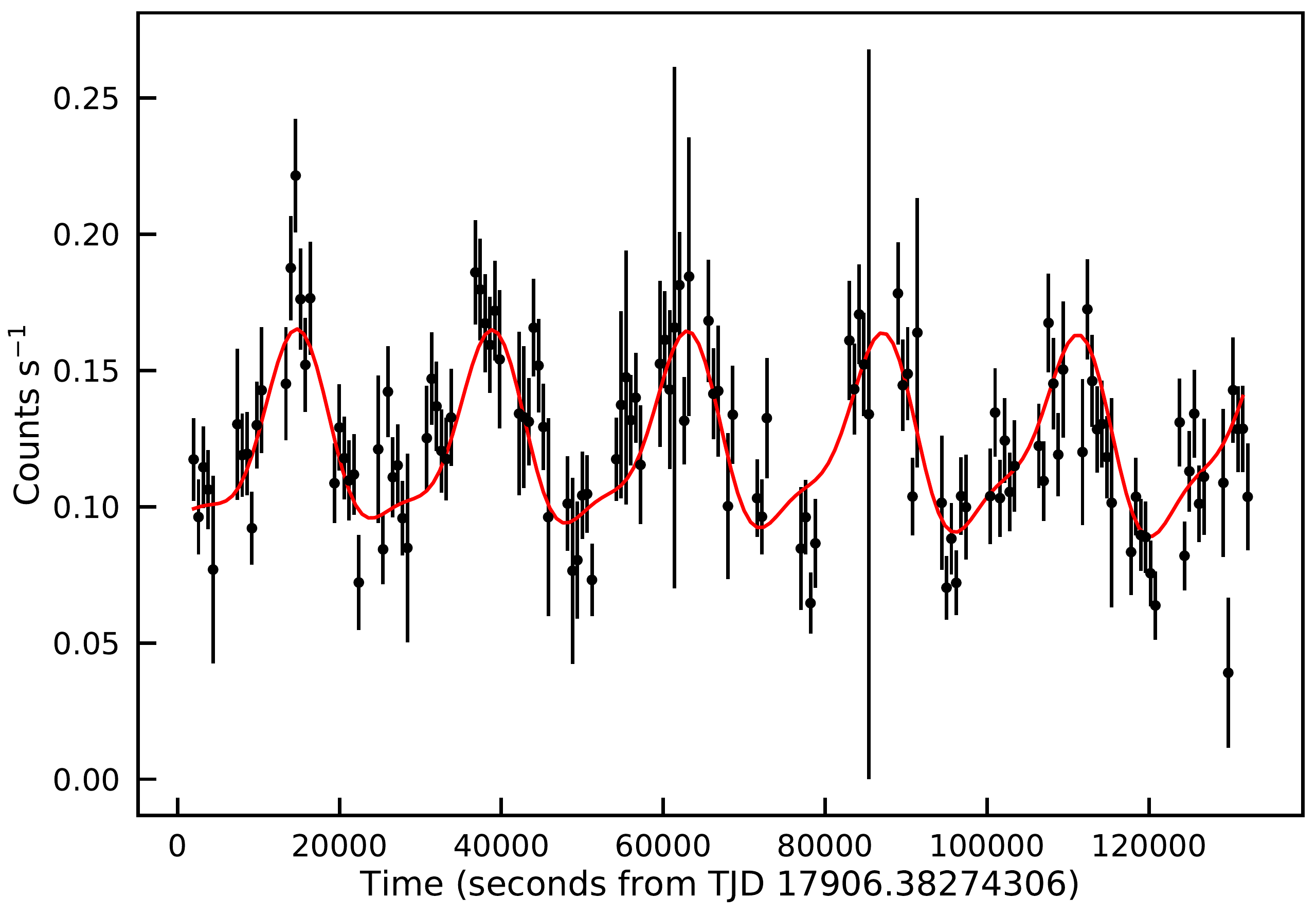}
\vspace{-7pt} 
\caption{Light curve of the combined FPMA and FPMB data in the 3 -- 15~keV energy range, binned at 600~s bin$^{-1}$. The red solid line marks the best fit, by assuming a model with two sinusoids plus a constant term.
}
\label{fig:lc2017}
%\vskip -0.1truecm
\end{center}
\end{figure}

The 3 -- 15~keV \nus\ events were used to study the timing properties
of the source. To obtain a rough
estimate of the period, we fitted the source light curve with a
constant plus two sinusoidal functions (see Figure
\ref{fig:lc2017}). The significance for the inclusion of the first 
harmonic was evaluated via the $F$-test. This yielded a probability 
of $\sim$ 5 $\times$ 10$^{-4}$ (3.6$\sigma$), showing that such a 
component was statistically required to improve the fit.
The best fit gave a fundamental period of 24061 $\pm$ 130~s (the first
harmonic period was forced to be half of the period). We refined this
result using an epoch folding search technique. We found a period $P$
= 24030 $\pm$ 40~s, where the error was evaluated as the 1 sigma
uncertainty on the width of the best-fitting Gaussian function used to
model the peak in the trial period distribution. The value is
compatible with the solution reported by
\citet{2011MNRAS.418..170E}.

\begin{figure}
\begin{center}
\includegraphics[scale=0.33]{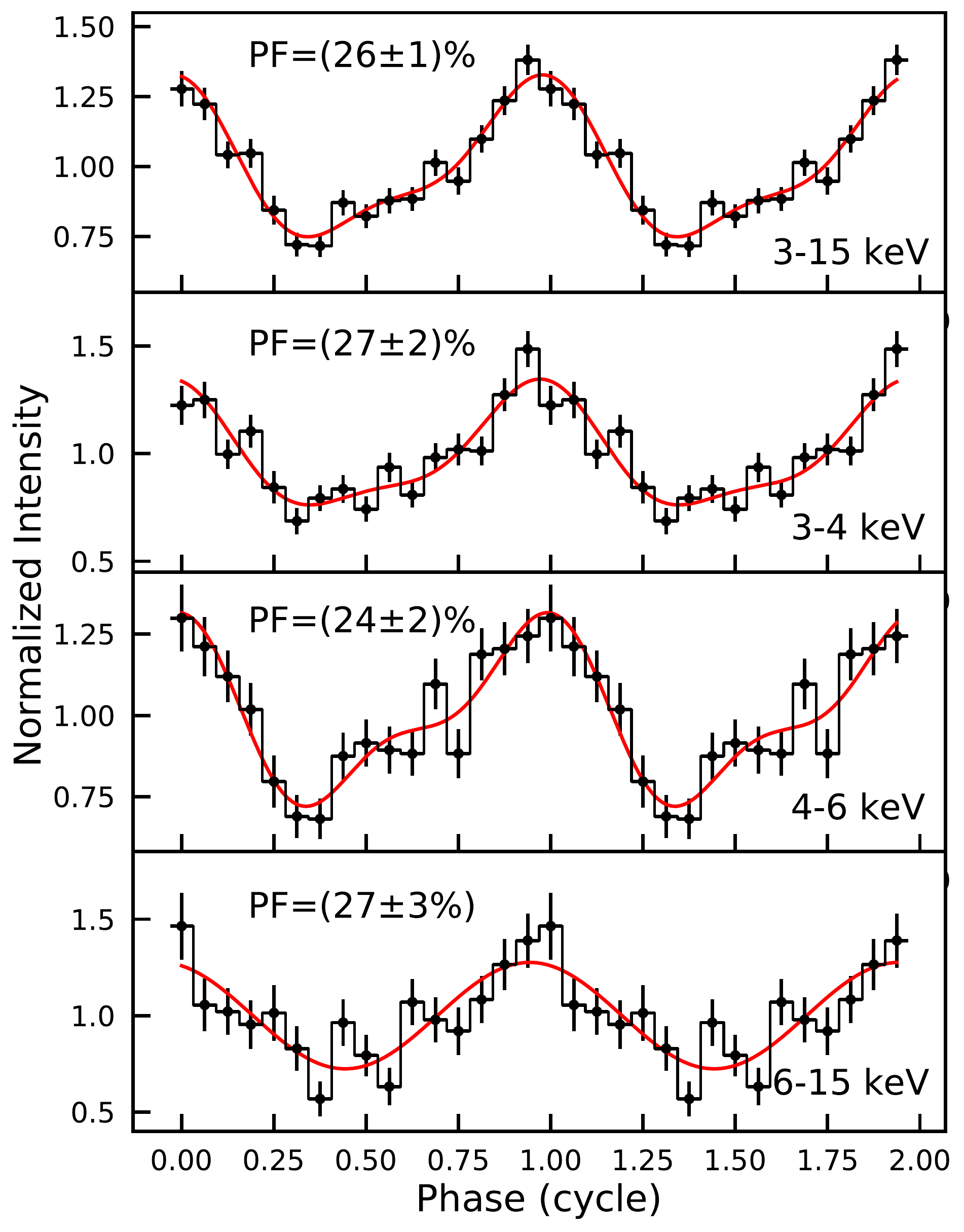}
\vspace{-7pt} 
\caption{16-bin pulse profiles in different energy bands, folded on $P$ = 24030~s and 57906 MJD as a reference epoch. The red solid lines define the best-fitting model (for more details see the text). The value of the pulsed fraction for each energy band is reported in the corresponding panel. Two cycles are shown for better visualization.}
\label{fig:pp2017}
%\vskip -0.1truecm
\end{center}
\end{figure}

We then folded the background-subtracted and exposure-corrected light
curves, sampled in 16 phase bins, on the best period $P$. We studied
the pulse profile in different energy bands, as shown in Figure
\ref{fig:pp2017}. The pulse profile changes shape as the energy
increases. In the 3 -- 4~keV and 4 -- 6~keV energy ranges two
sinusoidal functions are necessary to model properly the data
($F$-test probability of 0.003 and 3 $\times$ 10$^{-5}$,
respectively), while in the 6 -- 15~keV energy band only one sinusoid
is sufficient ($F$-test probability of 0.6 for the inclusion of the
first harmonic). The pulsed fraction, defined as the semi-amplitude of
the fundamental component divided by the source average count rate,
did not show any dependency on the energy, being consistent within the
errors in the different ranges considered and varying between 25 and
28 per cent without a clear tend. In the 3 -- 15~keV band the pulsed fraction inferred
value was (25.6 $\pm$ 1.2) per cent, a factor of $\sim$ 2 higher than
the value measured at the outburst onset, (14.0 $\pm$ 0.7) per cent.

Moreover, hardness ratios were computed considering different energy ranges with a comparable photon count statistics, and revealed no significant spectral variations along the rotational phase cycle.

%%%%%%%%%%%%%%%%%%%%%%%%%%%%%%%%%%%%%%%%%%%%%%%%%%%%%%%%%%%%%%%%%%%%%%%%%%%%%%%%%%%%%%%%%%%%
%%%%%%%%%%%%%%%%%%%%%%%%%%%%%%%%%%%%%%%%%%%%%%%%%%%%%%%%%%%%%%%%%%%%%%%%%%%%%%%%%%%%%%%%%%%

%%%%%%%%%%%%%%%%%%%%%%%%%%%%%%%%%%%%%%%%%%%%%%%%%%%%%%%%%%%%%%%%%%%%%%%%%%%%%%%%%%%%%%%%%%%%
%%%%%%%%%%%%%%%%%%%%%%%%%%%%%%%%%%%%%%%%%%%%%%%%%%%%%%%%%%%%%%%%%%%%%%%%%%%%%%%%%%%%%%%%%%%

\subsection{Spectral analysis}
\label{spec_ana}
All the background-subtracted spectra were grouped in order to have at
least 20 photon counts in each energy channel. The spectral modelling
was performed within the {\sc xspec} analysis package (version
12.9.1m, \citealt{1996ASPC..101...17A}) and using the $\chi^2$
statistics. To describe the absorption by the interstellar medium
along the line of sight, the {\sc tbabs} model was adopted with
photoionization cross-sections from \citet{1996ApJ...465..487V} and
chemical abundances from \citet*{2000ApJ...542..914W}. Once a best fit
was found, we applied the convolution model {\sc cflux} to calculate
the observed, unabsorbed fluxes and associated uncertainties.

\subsubsection{Phase-averaged spectral analysis}
\label{pha_ave}

To model the spectra in a wider energy range than the one covered by
\nus\ alone, we fit simultaneously the spectra for both
\nus\ FPMs together with a \swift\ XRT spectrum acquired $\sim$ 14
days later (ID: 00030389061). We restricted the energy range to 1 --
15~keV, where the source was not background dominated.

At the outburst peak (2016 June 25), the 1 -- 30~keV spectrum was well
described by a three-component model consisting of two blackbodies
with temperatures $kT_{BB_1}$ = 0.52 $\pm$ 0.01~keV and $kT_{BB_2}$ =
0.93 $\pm$ 0.05~keV, and inferred radii $R_{BB_1}$ = 2.7 $\pm$ 0.7~km
and $R_{BB_2}$ = 0.4 $\pm$ 0.2~km, with the addition of a power law
with photon index $\Gamma$ = 1.20 $\pm$ 0.25 to model the hard X-ray
tail \citep[2BB+PL hereafter;][]{2016ApJ...828L..13R}. As a first
step, we tried to fit the new spectra with the above-mentioned model,
freezing all the parameters at the outburst peak values and including
an overall constant to take into account the decay in flux. This
approach led to an unacceptable fit (reduced chi-square \rchisq = 2.34
for 190 degrees of freedom (dof)); residuals show an overestimation of
the emission at high energies ($\gtrsim$ 6~keV). Allowing all the parameters to vary,
we found the following results for the blackbody components:
$kT_{BB_1}$ = 0.54 $\pm$ 0.02~keV, $R_{BB_1}$ =
1.46$^{+0.21}_{-0.15}$~km, $kT_{BB_2}$ = 1.17$^{+0.11}_{-0.09}$~keV,
and $R_{BB_2}$ = 0.09$^{+0.03}_{-0.02}$~km; while the power law
component was not constrained (\rchisq = 1.08 for 184 dof). These
outcomes suggested that the spectral shape has simplified since the
outburst onset, without the need for a third spectral component to
model the data.

Therefore, as a next step, we tested the combination of two blackbodies (2BB) and the superposition of a blackboby and a power law (BB+PL). For all spectral fits, a renormalization factor was included to account for calibration uncertainties between different instruments and flux differences. We note that the XRT observation is much shorter than a phase cycle ($\sim$ 1/8) and, thus, samples a small interval of the rotational phase cycle. In view of the large flux modulation along the profile, XRT data could deviate from the averaged flux probed by \nus. For the XRT spectrum the factor was kept equal to 1, while for \nus\ spectra it was allowed to vary and was found to be consistent with 1 within the errors for both FPMs.

To better constrain the hydrogen column density \nh, we fit
simultaneously the new data sets with those acquired in 2016 at the
outburst peak \citep[see][for details about the data
  reduction]{2016ApJ...828L..13R}. This parameter was tied across all
the six spectra; its inferred value was (2.3 $\pm$ 0.1) $\times$
10$^{22}$~\cm\ for the 2BB model and (2.4 $\pm$ 0.1) $\times$
10$^{22}$~\cm\ for the BB+PL model.
The 2BB and BB+PL models provided a statistically equivalent fit with
a \rchisq = 1.09 for 861 dof. For the former, the best fit gave the
following results: $kT_{BB_1}$ = 0.51 $\pm$ 0.01~keV, $R_{BB_1}$ =
1.71$^{+0.15}_{-0.13}$~km, $kT_{BB_2}$ = 1.17$^{+0.07}_{-0.06}$~keV,
and $R_{BB_2}$ = 0.10 $\pm$ 0.02~km. We note that the hotter blackbody
has a temperature higher than that at the outburst onset and the
corresponding hotspot had shrunk of a factor of 4. This component is
most likely mimicking residual emission at higher energy; Figure
\ref{fig:sp2017}, left panel, shows the spectra fitted with this model
and the post-fit residuals. Adopting the BB+PL model, the spectra are
described by a blackbody with $kT_{BB_1}$ = 0.57 $\pm$ 0.01~keV and
$R_{BB_1}$ = 1.07$^{+0.08}_{-0.07}$~km, plus a soft power law with
$\Gamma$ = 3.9 $\pm$ 0.1 and normalization of (1.7 $\pm$ 0.4) $\times$
10$^{-2}$. The 0.5-30~keV observed flux was (9.3 $\pm$ 0.3) $\times$
10$^{-12}$~\flux, with a fractional contribution of 98\% from the 0.5
-- 10~keV energy band. The corresponding unabsorbed flux was (1.99
$\pm$ 0.08) $\times$ 10$^{-11}$~\flux\, which translates into a
luminosity of (2.6 $\pm$ 0.1) $\times$ 10$^{34}$~\lum. These results
are indicative of an anisotropic distribution of the surface
temperature, as already observed in other magnetars and theoretically
predicted \citep{2013MNRAS.434.2362P}. For the 2016 data sets we
retained the 2BB+PL model and obtained results consistent within the
errors with those presented in our previous work for the two blackbody
components, whereas we got a slightly steeper power law with a photon
index $\Gamma$ = 1.8 $\pm$ 0.2 and a normalization of
(5.7$_{-2.4}^{+4.2}$) $\times$ 10$^{-4}$. Figure \ref{fig:sp2017},
right panel, shows the unfolded spectra relative to the latest
observations and to the outburst onset, fitted simultaneously, and the
residuals with respect to the BB+PL model for the former and the
2BB+PL model for the latter.

In order to check if the random sampling of the \swift\ data along the phase cycle could affect the spectral fitting results, we performed again the analysis including the XRT spectrum collected $\sim$ 15 days before the \nus\ data (ID: 00030389060). Also this XRT observation has an exposure time equal to $\sim$ 1/8 of the rotational period. The values of the spectral parameters are compatible within the errors with those reported above. Moreover we verified that the spectral modelling results were not affected by the size and location of the background extraction region. Table \ref{tab:fit} summarises the spectral fitting results; for a discussion about the physical interpretations of the adopted models see Section \ref{disc}.\\

\begin{figure*}
\begin{center}
\includegraphics[scale=0.33]{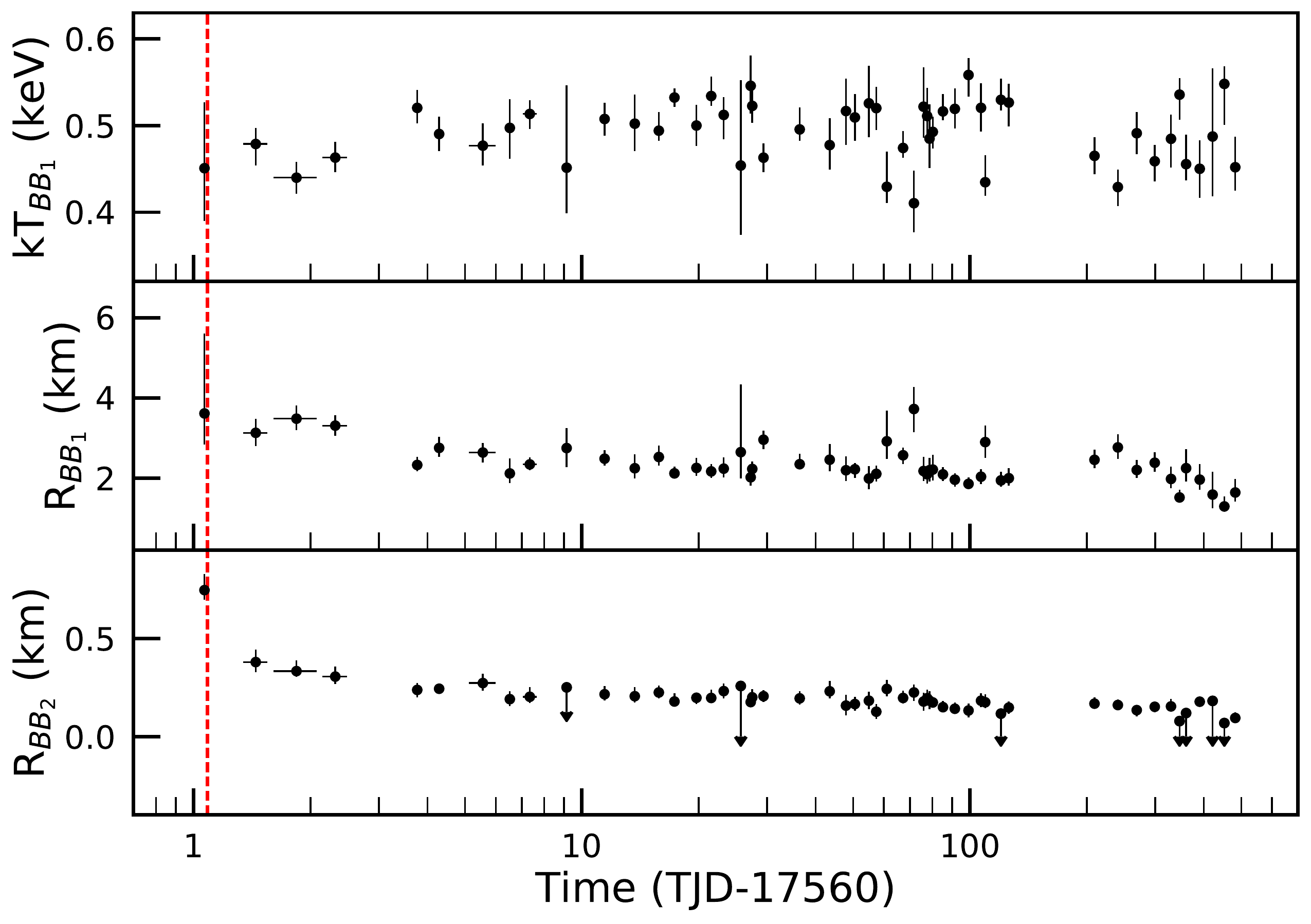}
\includegraphics[scale=0.33]{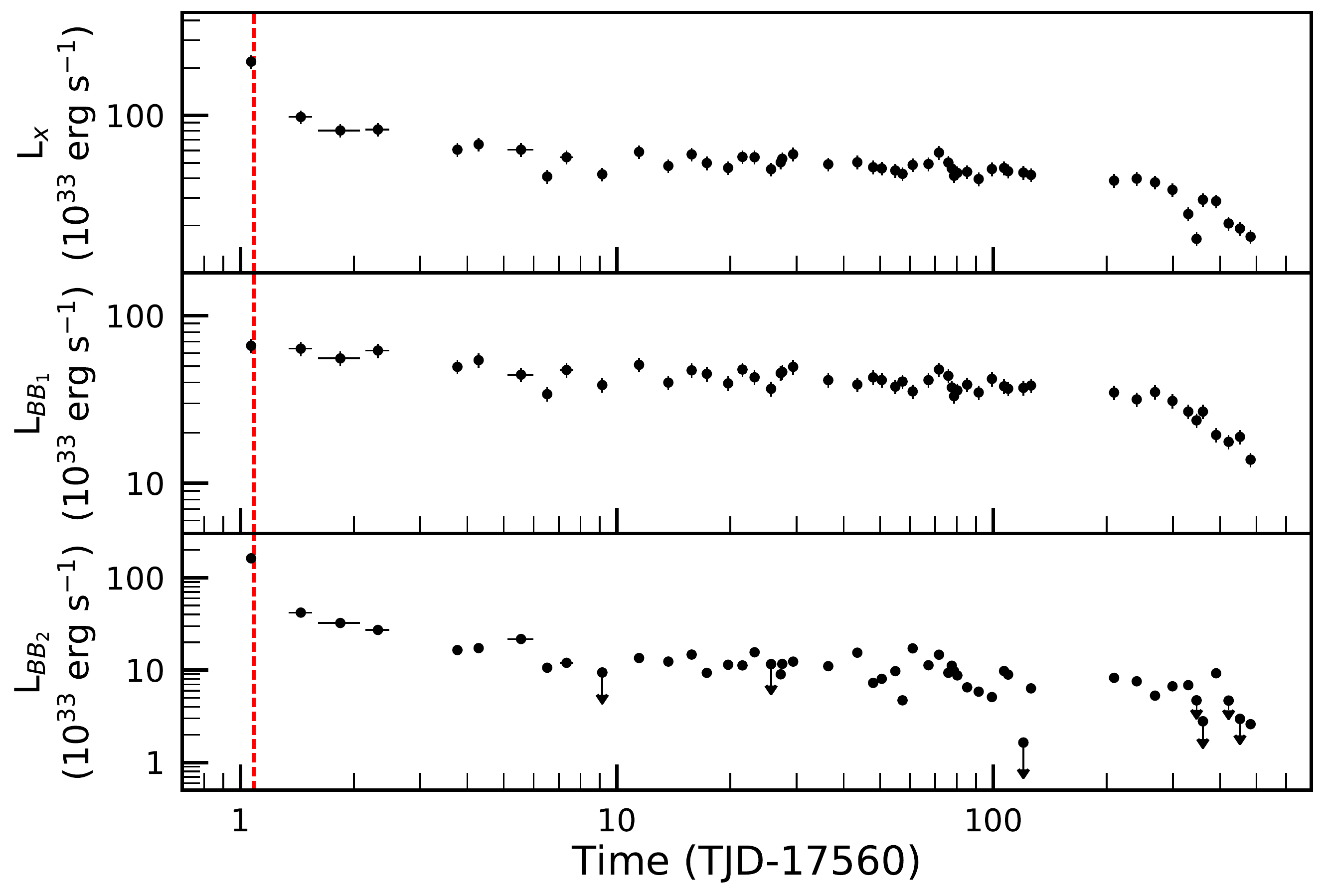}
\vspace{-5pt} 
\caption{
Time evolution of the blackbody temperature for the colder component and radii for both blackbody components (left panel), luminosities of the total emission and of the single components in the 0.5 -- 10~keV energy band (right panel) for the \swift\ XRT data of the monitoring campaign. The red vertical dashed line corresponds to the epoch of the \swift\ BAT trigger on 2016 June 22 at 02:03:13 UT \citep{2016ATel.9180....1D}.   
}

\label{fig:monitoringpar}
%\vskip -0.1truecm
\end{center}
\end{figure*}

\begin{figure*}
\begin{center}
\includegraphics[scale=0.33]{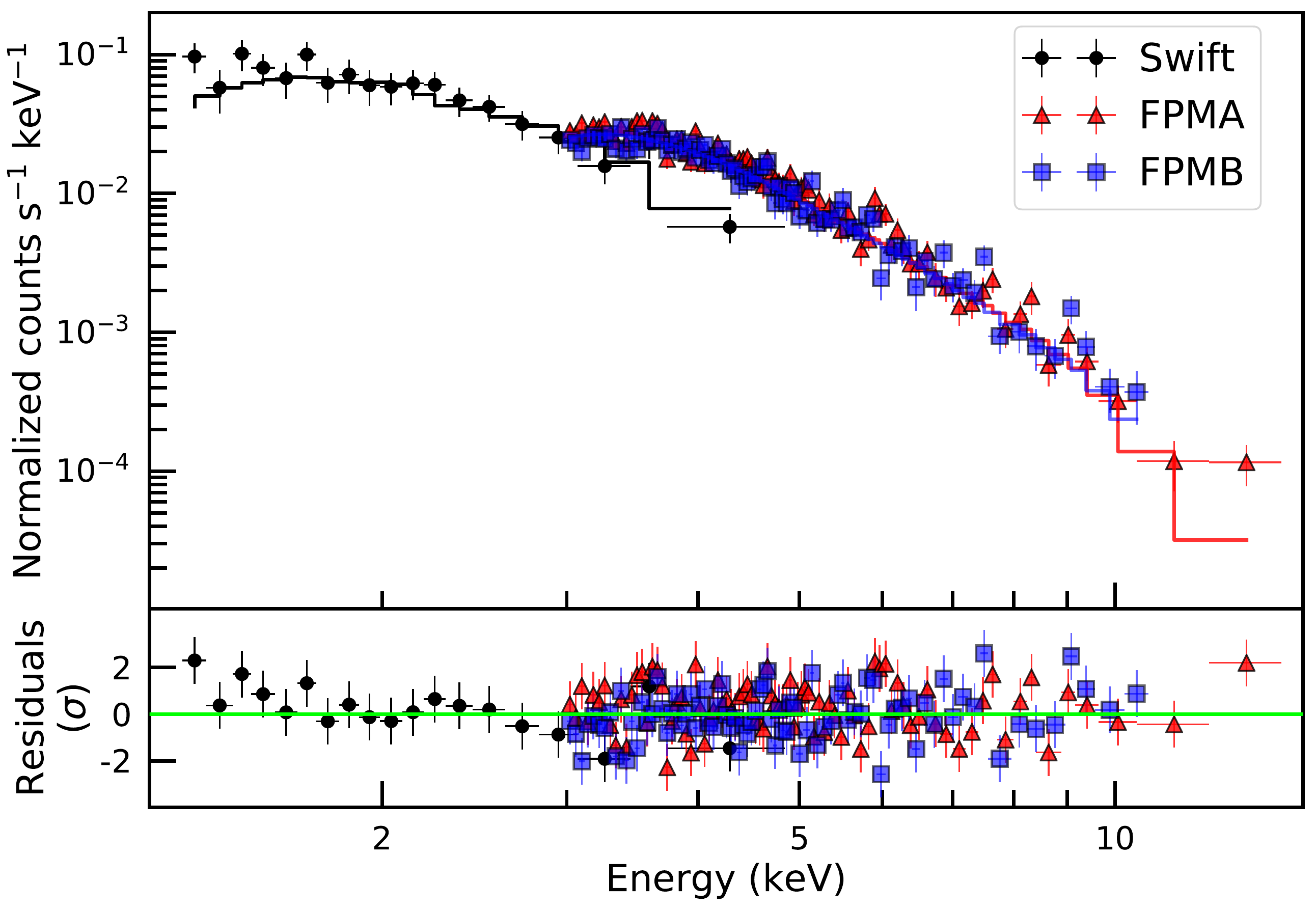}
\includegraphics[scale=0.33]{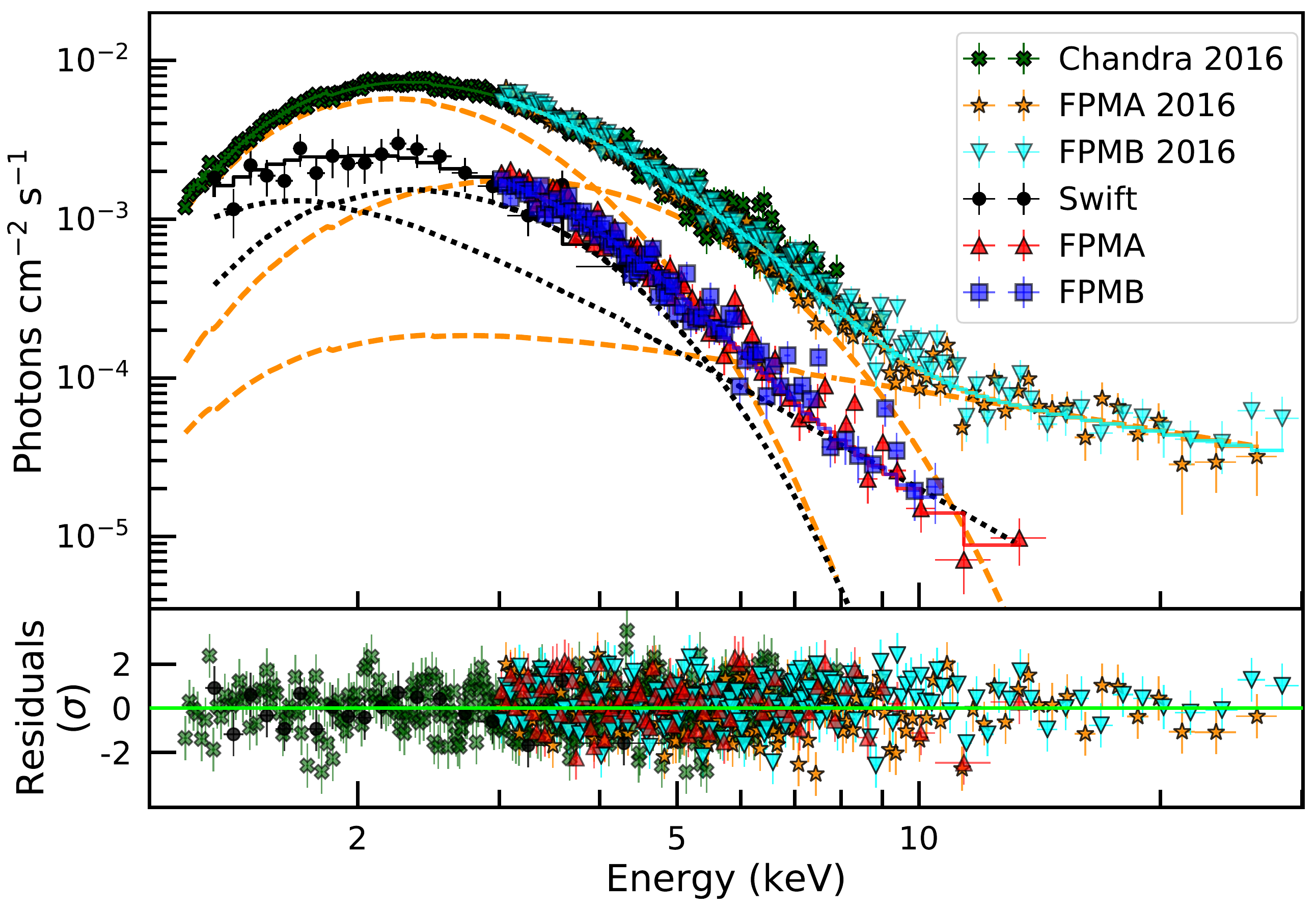}
\vspace{-7pt} 
\caption{ Left panel: phase-average spectra relative to the 2017
  \swift+\nus\ observations. The solid line represents the model
  consisting of two absorbed blackbodies. Residuals with respect to
  this model are shown in units of standard deviations in the bottom
  panel. Right panel: unfolded spectra for the 2016 and 2017 data sets,
  fitted with an absorbed double blackbody plus a power law model and
  with the superposition of an absorbed blackbody and a power law
  component, respectively (see text for more details). Orange dashed (2016)
  and black dotted (2017) lines represent the different components of the
  models. Post-fit residuals in units of standard deviations are shown
  in the bottom panel.}

\label{fig:sp2017}
%\vskip -0.1truecm
\end{center}
\end{figure*}

While the individual \swift\ observations do not have the photon
statistics necessary for a detailed spectral analysis, they are useful
to study the luminosity evolution of the source. We restricted the
spectral modelling to the 1 -- 10~keV energy band and adopted a
two-component model, without including the hard power law discovered
at the outburst peak. This non-thermal component is a transient
feature of the source spectrum and its contribution to the source flux
was significant above 10~keV, i.e. outside the energy interval covered
by XRT. We fit all the spectra together, fixing the absorption value
to be the same among all of them. We chose an absorbed double
blackbody model in order to follow the evolution of the hotter
component, that is possibly linked to the outburst mechanism. The
effective temperature for the hotter blackbody was initially left free
to vary across the data sets; however it was found to be poorly
constrained in the single exposures. Therefore we decided to tie the
hotter blackbody temperature across all the spectra, to 
better constrain the time evolution of the size of the emitting region for the
hotter component. We obtained an acceptable description of the data (
$\chi^2_\nu = 0.95$ for 2511 dof), with $N_H=2.25_{-0.06}^{+0.05}
\times10^{22}$~cm$^{-2}$, $kT_{BB_2}=1.24_{-0.07}^{+0.08}$~keV. We
measured a shrinking of the hotter blackbody emitting area: the
inferred radius decreased from $\sim$ 0.7~km, measured at the onset,
to $\sim$ 0.1~km, about 480 days later. The time evolution of the
best-fitting model parameters, the luminosities of the single
components and of the total emission (all in the 0.5 -- 10~keV energy
band) are shown in Figure \ref{fig:monitoringpar}.
  
%%%%%%%%%%%%%%%%%%%%%%%%%%%%%%%%%%%%%%%
\begin{table*}
\caption{Results of the simultaneous spectral fitting for the \nus\ and \swift\ observations of \src\ performed on 2-16 June 2017. The X-ray fluxes and luminosities are calculated in the 0.5-30~keV energy range. All errors are quoted at 1$\sigma$ confidence level.}
\label{tab:fit}

\small{
\begin{tabular}{@{}lcccccccc}
\hline
Model & \nh & $kT_{BB_1}$  & $R_{BB_1}$ & $kT_{BB_2}$/$\Gamma$ & $R_{BB_2}$/PL norm$^a$ &  $F_{X,abs}$  & $L_X$  & \rchisq (dof) \\ 
	  & (10$^{22}$~\cm) & (keV) &  (km) &  (keV)/- & (km)/(10$^{-2}$)  & (10$^{-12}$~\flux) & (10$^{34}$~\lum) &               \\
\hline
2BB   & 2.3 $\pm$ 0.1 & 0.51 $\pm$ 0.01 & 1.71$^{+0.15}_{-0.13}$ & 1.17$^{+0.07}_{-0.06}$ & 0.10 $\pm$ 0.02 & 9.1 $\pm$ 0.1 & 2.5 $\pm$ 0.1 & 1.09 (861) \\
\hline
BB+PL & 2.4 $\pm$ 0.1 & 0.57 $\pm$ 0.01 & 1.07$^{+0.08}_{-0.07}$ & 3.9 $\pm$ 0.1 & 1.7 $\pm$ 0.4 & 9.3 $\pm$ 0.3 & 2.6 $\pm$ 0.1 & 1.09 (861) \\
\hline
\end{tabular}}
\begin{list}{}{}
\item[$^{a}$]The power law normalization is in units of photons/keV/cm$^2$/s at 1~keV.
\end{list}

\end{table*}
%%%%%%%%%%%%%%%%%%%%%%%%%%%%%%%%%%%%%%%

%%%%%%%%%%%%%%%%%%%%%%%%%%%%%%%%%%%%%%%%%%%%%%%%%%%%%%%%%%%%%%%%%%%%%%%%%%%%%%%%%%%%%%%%%%%
%%%%%%%%%%%%%%%%%%%%%%%%%%%%%%%%%%%%%%%%%%%%%%%%%%%%%%%%%%%%%%%%%%%%%%%%%%%%%%%%%%%%%%%%%%

%%%%%%%%%%%%%%%%%%%%%%%%%%%%%%%%%%%%%%%%%%%%%%%%%%%%%%%%%%%%%%%%%%%%%%%%%%%%%%%%%%%%%%%%%%%%
%%%%%%%%%%%%%%%%%%%%%%%%%%%%%%%%%%%%%%%%%%%%%%%%%%%%%%%%%%%%%%%%%%%%%%%%%%%%%%%%%%%%%%%%%%%

\subsubsection{Phase-resolved spectral analysis }

To study the spectral variation as a function of the rotational phase,
we produced a normalized energy versus phase image for the combined
FPMA and FPMB event file, binning the source counts in 100 rotational
phase bins and 150 energy channels, and then normalizing it to the
pulse profile and phase-averaged spectrum. The energy range was
narrowed to 3 -- 15~keV, where the source was detected, and we used
our best period ($P$ = 24030~s) to calculate the phase. This method
allowed us to investigate the presence of spectral feature that vary
with the phase and/or energy (as observed in the magnetar SGR
0418+5729; \citealt{2013Natur.500..312T}), without assuming a specific
spectral energy distribution. No significant features were detected in
the normalized image; we also tried different binnings and found the
same result.

\begin{figure*}
\begin{center}
\includegraphics[scale=0.33]{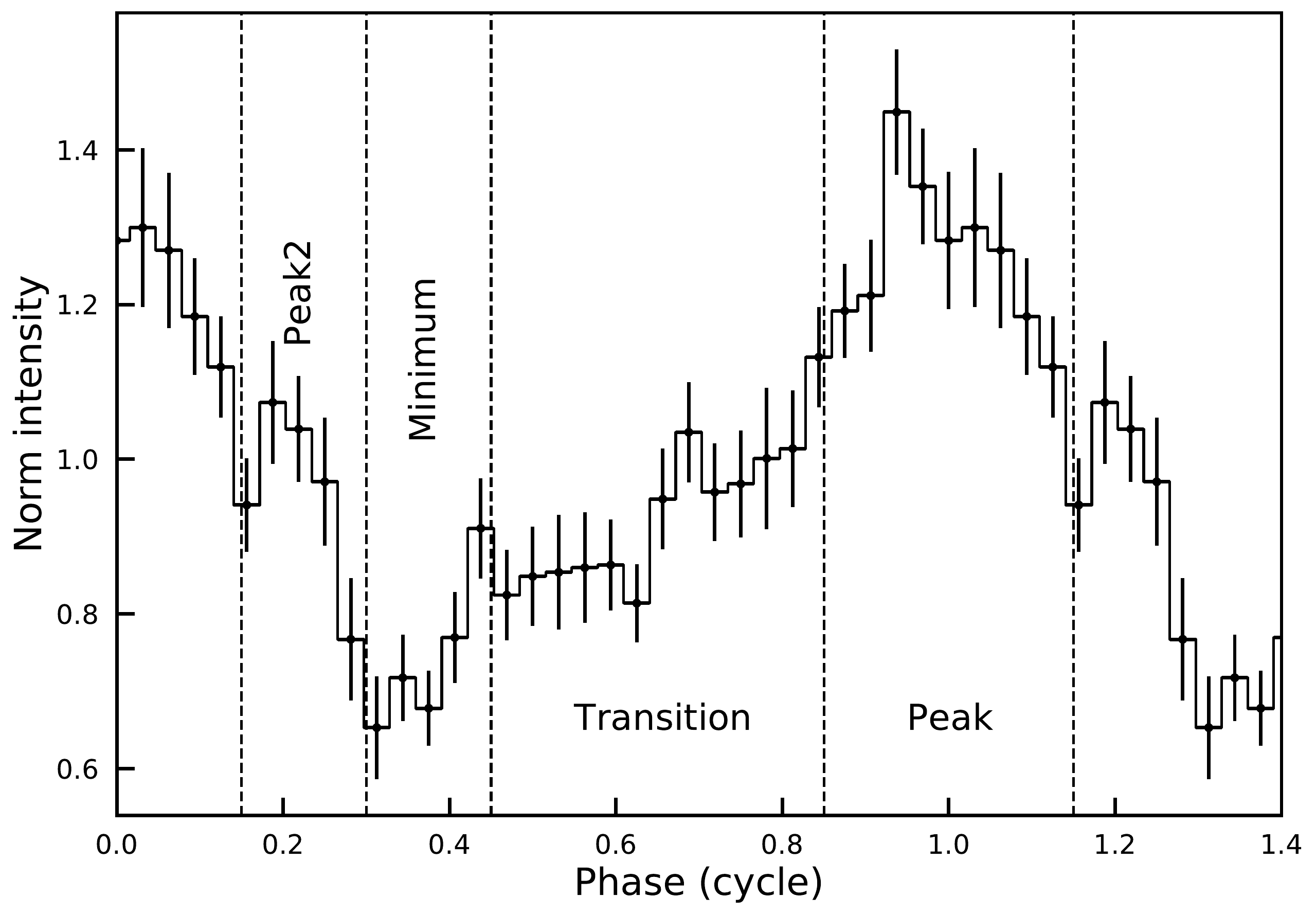}
\includegraphics[scale=0.33]{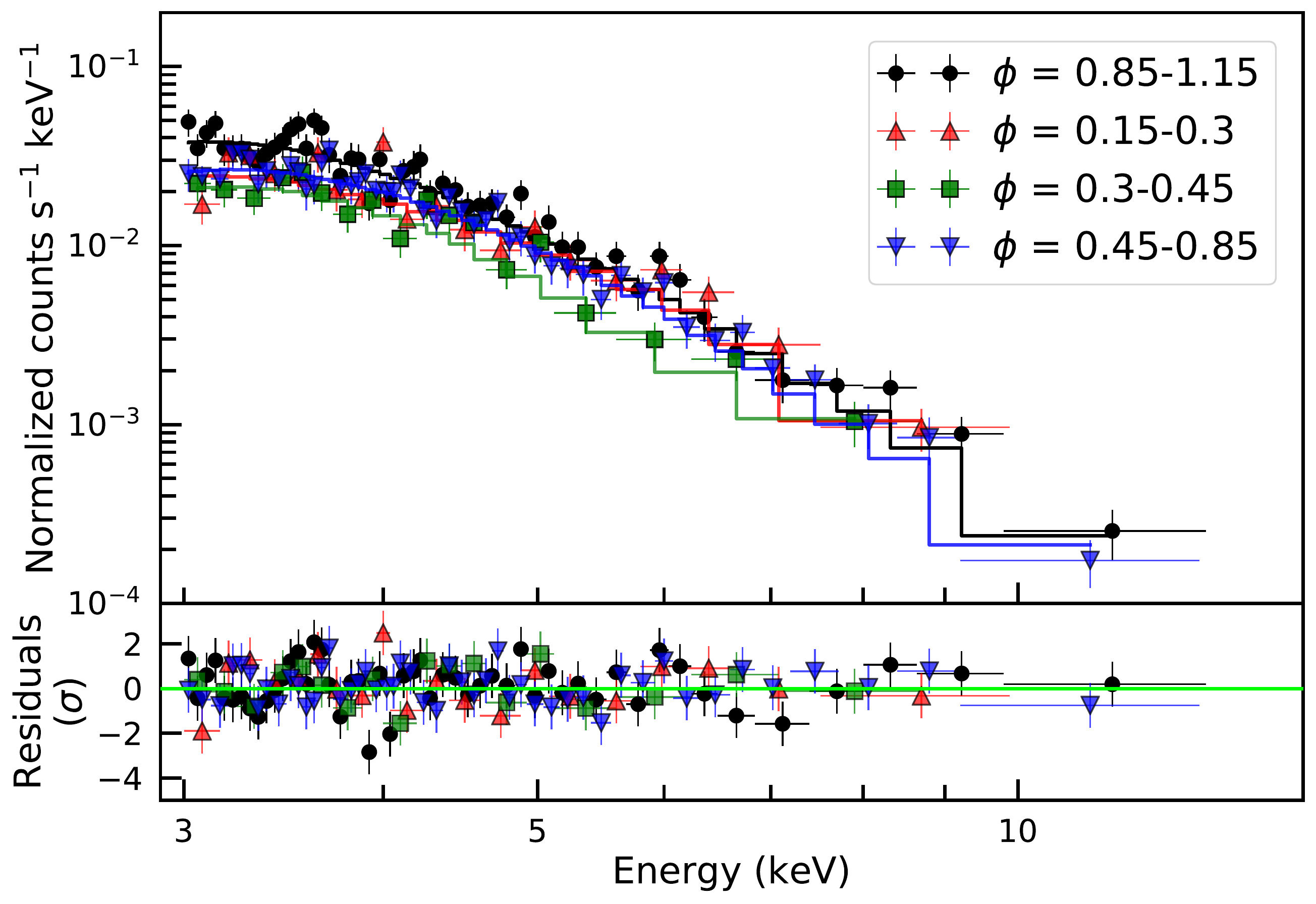}
\vspace{-5pt} 
\caption{ Left panel: Pulse profile sampled with 32 phase bins in the 3 -- 15~keV energy band. The
  period and the reference epoch are the same as in Figure
  \ref{fig:pp2017}. The intervals used for the phase-resolved
  spectroscopy are marked with dashed lines. Right panel:
  phase-resolved spectra extracted from \nus\ FPMA data fitted
  simultaneously. The solid lines denote the model consisting of an
  absorbed blackbody with the inclusion of a power law. Post-fit
  residuals in units of standard deviations are also shown.  }

\label{fig:pps}
%\vskip -0.1truecm
\end{center}
\end{figure*}

We then carried out a phase-resolved spectroscopy for \nus\ FMPA data to analyse distinct features in the pulse profile. We divided the phase cycle in four intervals: 0.3 -- 0.45 (corresponding to the minimum of the modulation), 0.45 -- 0.85 (referred to as a transition region), 0.85 -- 1.15 (related to the main peak) and 0.15 -- 0.3 (where a second, less prominent peak of marginally significance is present; see Figure \ref{fig:pps}, left panel).

The spectral analysis was performed assuming the BB+PL model in the energy range 3 -- 15~keV and the \nh\ was fixed to the phase-averaged value (\nh\ = 2.4 $\times$ 10$^{22}$~\cm). Firstly, we only left the normalizations of each component free to vary, while the blackbody temperature and photon index were fixed to their best fit values for the phase-averaged spectrum. This fit gave a satisfactory description of the spectra with a \rchisq = 0.94 (130 dof); we note that the size of the blackbody emitting area increases along the phase cycle, while the normalization of the power law is consistent within the error for the different phase-resolved spectra. Allowing all the parameters to vary among the spectra yielded a statistically equivalent fit (\rchisq = 0.93 for 130 dof). The spectral parameters do not show any significant variability as a function of the $\sim$ 24-ks periodicity.

%%%%%%%%%%%%%%%%%%%%%%%%%%%%%%%%%%%%%%%%%%%%%%%%%%%%%%%%%%%%%%%%%%%%%%%%%%%%%%%%%%%%%%%%%%%
%%%%%%%%%%%%%%%%%%%%%%%%%%%%%%%%%%%%%%%%%%%%%%%%%%%%%%%%%%%%%%%%%%%%%%%%%%%%%%%%%%%%%%%%%%

\section{Discussion}
\label{disc}

We reported on a new \nus\ observation (the second performed so far in the hard X-rays) of the compact object at the center of the SNR \snr, \src,  performed 345 days after the onset of its last outburst, as well as on a \swift\ XRT monitoring campaign that covered a timespan of $\sim$ 480 days of the outburst decay. The monitoring campaign resumed at the beginning of 2018 with one observation per month (PI: De Luca); the updated results will be available at the Magnetar Outburst Online Catalogue\footnote{\url{http://magnetars.ice.csic.es}} \citep{2018MNRAS.474..961C}.

About one year after the last flaring event, the source is above the
background level up to $\sim$ 15~keV. The X-ray spectrum can be well
reproduced by an absorbed blackbody with the addition of another
component at higher energy (either a second blackbody or a power law),
with no need for an additional power-law like component (as detected
up to $\sim$ 30~keV at the outburst peak; Rea et al. 2016). This
clearly indicates a softening of the source emission with time. The
colder blackbody component attains a relatively constant temperature
$kT_{BB_1}$ $\sim$ 0.5 -- 0.6~keV during the outburst decay, whereas
the corresponding emitting region is shrinking. The inferred 0.5 --
10~keV observed flux at the epoch of the \nus\ observation was $\sim$
9 $\times$ 10$^{-12}$~\flux, about one order of magnitude higher than
the historical quiescent level measured by \cxo\ in 1999
\citep{2000HEAD....5.3211G} and smaller by a factor of $\sim$ 3 than
what was observed at the outburst onset, in June 2016.

\subsection{Comparison with magnetars}

The superposition of two blackbodies or a blackbody plus a power law is usually applied to describe soft X-ray emission from magnetars \cite[see][Table 2]{2018MNRAS.474..961C}. In magnetars, this is generally interpreted as the thermal emission associated with the cooling of the NS surface that could be distorted by resonant cyclotron scattering of thermal photons from the surface onto the charged particles flowing in a twisted magnetosphere. Thermal photons produced at the surface gain energy through repeated scatterings with charged particles that flow along the magnetic field lines, leading to the formation of a tail at higher energy, that often is modelled with a power law. 

Considering the similarities between the members of the magnetar class
and this source, we performed a fit of the most recent \nus\ data with
a more physical model, the NTZ model, that accounts for resonant
cyclotron up-scattering of the soft seed photons
\citep*{2008MNRAS.386.1527N,2008MNRAS.389..989N}. This model assumes a
uniformly heated NS surface and a globally twisted magnetosphere. We
obtained a satisfactory description of the data (\rchisq = 1.08 for
186 dof) by leaving all the parameters free to vary, deriving
constraints for the bulk velocity of the charged particles in the
magnetosphere ($\beta_{bulk}$ = 0.205$^{+0.008}_{-0.006}$) and the
twist angle ($\Delta \phi$ = 0.48 $\pm$ 0.02~rad), while the values
inferred for the hydrogen column density and the surface temperature
are consistent with the results presented in Section
\ref{pha_ave}. For comparison we also fit the NTZ model to the data
sets acquired at the outburst onset. In this case, an additional power
law component was required to model the hard non-thermal tail. We
obtained an acceptable modelling of the spectra (\rchisq = 1.14 for
676 dof). The bulk velocity was consistent with the previous
measurement, while the twist angle was 0.72 $\pm$ 0.02~rad. The fact
that the NTZ model provides a good fit at both epochs is an indication
of the presence of some magnetospheric distortion (e.g. twisted
magnetic field lines). Although some caveats should be adopted in
interpreting the time evolution of the parameters inferred from the
NTZ model (as it assumes a global twist of the magnetosphere), the
decrease in the twist angle suggests a gradual untwisting of the star
magnetosphere, as expected for magnetars while recovering from their
outburst.
Furthering the magnetar analogy, the detection of a small hot spot on
the NS surface at the outburst peak \citep{2016ApJ...828L..13R} is
suggestive of a scenario in which the magnetic twist is localized in a
limited area of the magnetosphere, most likely in the form of a
current-carrying bundle of field lines
\citep{2009ApJ...703.1044B}. Once formed due to crust
displacements (triggered by internal magnetic stresses, \citealt{1995MNRAS.275..255T,2011ApJ...727L..51P}),
the bundle has to decay to support its own currents and its gradual dissipation
leads to a reduction of the region at the footprints of the
twist. This interpretation is in agreement with the shrinking of the
hotter thermal component, qualitatively observed thanks to the
\swift\ monitoring campaign. Furthermore, when the bundle untwists,
the charged particle density decreases, as well as the scattering
optical depth, making the resonant cyclotron scattering less
efficient. This implies that the hard X-ray tail filled up by
up-scattered thermal photons becomes less populated, producing an
overall softening of the spectrum during the outburst decay. This
behaviour is commonly observed in other magnetar outburst decays
\citep{2011ASSP...21..247R,2018MNRAS.474..961C}. \\

The prolonged \swift\ XRT monitoring allowed us to refine the empirical modelling of the outburst decay light curve. Following \citet{2016ApJ...828L..13R} and \citet{2018MNRAS.474..961C}, we fitted to the data points a double exponential function plus a constant term of the form: 
\begin{equation}
L(t)=L_q + \sum_{i=1}^2 A_i\times \rm{exp}(-$$t$$/\tau_i)~,
\end{equation} 
where $\tau_i$ represents the $e$-folding time and can be used as an estimate of the time scale of the decay. In the fit, $L_q$ was
fixed to the quiescent value measured with \cxo\ in 1999 September,
$L_q \sim 2.8\times10^{33}$~\lum. Best fitting parameters were $\tau_1
= 0.44_{-0.09}^{+0.14}$~d, $\tau_2 = 406_{-19}^{+20}$~d, $A_1 =
(3.6\pm0.8) \times10^{34}$~\lum, $A_2 = (5.2\pm0.1)
\times10^{34}$~\lum. We consider $\tau_2\approx400$\,d as the fundamental time scale of the outburst decay, as it traces the long-term evolution of the light curve after the faster but short-lived (on a characteristic time scale $\tau_1\approx0.4$\,d ) flux decrease observed immediately after the onset of the outburst.
Extrapolation of this phenomenological model
up to the approximate epoch of return to the quiescent state
($\sim$ 2021, as predicted by the model) leads to an estimated released energy
of about $2\times10^{42}$~erg. This is only slightly lower than the
value computed using a smaller number of data points
\citep{2018MNRAS.474..961C}.
%We use $\tau_2$ to estimate the time scale of the outburst decay, as it traces the long-term evolution of the light curve following the initial rapid decay observed over the first few days since the outburst onset.

\src\ clearly follows several different correlations observed in
magnetars (see Fig.~3, 6 and 8 by \citealt{2018MNRAS.474..961C}). For
example, it follows the anti-correlation between the quiescent X-ray
luminosity and the outburst luminosity increase. This relation
strongly suggests that magnetars in outburst cannot exceed a
luminosity of $\sim 10^{36}$~\lum\ at the peak; \src\ reached a
maximum luminosity of $\sim 2 \times 10^{35}$~\lum\ in its last
outburst, a factor of $\sim 100$ above the quiescent X-ray luminosity. For magnetars a correlation between the total energy emitted during the outburst and the maximum luminosity reached at the outburst onset is significant at 4$\sigma$, implying that the more energetic outbursts reach a larger peak luminosity. Moreover, the total outburst energy correlates with the time scale of the decay, meaning that the longer the outburst, the more energetic.
%Furthermore for magnetars the energetic correlates with the maximum
%X-ray luminosity reached at the peak of the outburst and with the
%decay timescale, implying that the more energetic outbursts reach a
%larger luminosity at the peak, and the longer the outburst, the more
%energetic. 
The above-mentioned updated and refined values for the
total energy and the decay time scale further corroborate a
(phenomenological) classification of \src\ as a magnetar. In fact,
although \src\ has not reached its quiescent level yet, the energy
released of 2 $\times$ 10$^{42}$~erg and the decay time scale of $\sim$
400~d are so far in agreement with what is expected from magnetars,
following the above-mentioned correlation. We point out that
\src\ represents the only case so far that allows us to highlight such
a phenomenological link between the members of the CCO class and
magnetars, since all other known CCOs are steady X-ray emitters. Future
detections of outbursts from known CCOs might help strengthen
such a connection.

\section*{Acknowledgements}
The results reported in this article are based on observations obtained with \swift\ and \nus. \swift\ is a NASA mission with participation of the  Italian Space Agency and the UK Space Agency. The \nus\ mission is a project led by the Californian Institute of Technology. AB, PE and NR are supported by an NWO Vidi Grant (PI: Rea). FCZ and NR are supported by grants AYA2015-71042-P and SGR 2014-1073. We thank the PHAROS COST Action (CA16214) for partial support and the referee for the comments.

\bibliographystyle{mnras}
\bibliography{reference}

\end{document}